# Compressed H$_3$S: inter-sublattice Coulomb coupling in a high-$T_C$ superconductor*

**Dale R. Harshman**[1] and **Anthony T. Fiory**[2]

[1] Department of Physics, The College of William and Mary, Williamsburg, VA 23187, USA
[2] Department of Physics, New Jersey Institute of Technology, Newark, NJ 07102, USA

E-mail: drh@physikon.net

**Abstract**

Upon thermal annealing at or above room temperature (RT) and at high hydrostatic pressure $P \sim 155$ GPa, sulfur trihydride H$_3$S exhibits a measured maximum superconducting transition temperature $T_C \sim 200$ K. Various theoretical frameworks incorporating strong electron-phonon coupling and Coulomb repulsion have reproduced this record-level $T_C$. Of particular relevance is that experimentally observed H-D isotopic correlations among $T_C$, $P$, and annealed order indicate an H-D isotope effect exponent α limited to values ≤ 0.183, leaving open for consideration unconventional high-$T_C$ superconductivity with electronic-based enhancements. The work presented herein examines Coulombic pairing arising from interactions between neighboring S and H species on separate interlaced sublattices constituting H$_3$S in the Im$\bar{3}$m structure. The optimal value of the transition temperature is calculated from $T_{C0} = k_B^{-1} \Lambda e^2 / \ell \zeta$, with $\Lambda = 0.007465$ Å, inter-sublattice S-H separation spacing $\zeta = a_0 / \sqrt{2}$, interaction charge linear spacing $\ell = a_0 (3/\sigma)^{1/2}$, average participating charge fraction $\sigma = 3.43 \pm 0.10$ estimated from calculated H-projected electron states, and lattice parameter $a_0 = 3.0823$ Å at $P = 155$ GPa. The resulting value of $T_{C0} = 198.5 \pm 3.0$ K is in excellent agreement with transition temperatures determined from resistivity (196 – 200 K onsets, 190 – 197 K midpoints), susceptibility (200 K onset), and critical magnetic fields (203.5 K by extrapolation). Analysis of mid-infrared reflectivity data confirms the expected correlation between boson energy and $\zeta^{-1}$. Suppression of $T_C$ below $T_{C0}$, correlating with increasing residual resistance for < RT annealing, is treated in terms of scattering-induced pair breaking. Correspondences between H$_3$S and layered high-$T_C$ superconductor structures are also discussed, and a model considering Compton scattering of virtual photons of energies ≤ $e^2/\zeta$ by inter-sublattice electrons is introduced, illustrating that $\Lambda \propto \lambda_C$, where $\lambda_C$ is the reduced electron Compton wavelength.

---

# 1. Introduction

The highest known superconducting transition temperature $T_C$ of 203 K (onset) has been reported for a highly compressed phase of $H_3S$ synthesized from $H_2S$ and formed at pressure $P$ near 155 GPa [1]. The stable high-$T_C$ phase of $H_3S$ at high $P$ is presumed to have $Im\bar{3}m$ symmetry, while lower-$T_C$ phases of R3m structure exist at reduced $P$ [2-7].[3] Rhombohedral distortion for $P < 140$ GPa associated with R3m symmetry has been reported for $H_3S$ formed by reacting $H_2$ and S [7]. Considering theoretical results of S-H bonding in the $Im\bar{3}m$ structure, the nearest-neighbor bonds along the principle cube axes are strongly covalent and metallic, whereas the next-nearest-neighbors of S-H are weakly bonded [2]. Thus, the $Im\bar{3}m$ structure can be viewed as comprising two interlaced simple cubic $H_3S$ metallic sublattices with essentially non-bonding interactions between them [8].

Annealing samples to reduce disorder is evidently crucial for observing high $T_C$ in $H_3S$. For example, the onset temperature 203(1) K reported in [1] is obtained from temperature extrapolation of critical magnetic field data on a sample annealed at or above room temperature (RT), even as the bulk of the sample remained electrically resistive above 170 K. Absent such annealing, the observed transition temperatures are dramatically reduced, e.g. from measurements taken over temperature ranges up to 260 K, $T_C \sim 65 - 150$ K for $P \sim 155 - 200$ GPa [1]. These depressed values of $T_C$, attributed to annealing kinetics, correlate with higher normal-state resistances and larger superconducting transition widths [1]. Maxima observed in the pressure dependences of $T_C$ in $H_3S$ and $D_3S$, and also the depressed $T_C(P)$ for $D_3S$ relative to $H_3S$, were previously shown to correlate with residual electrical resistance ratios and indicate variations in sample quality [9].

Calculations based on strong electron-phonon coupling and Eliashberg theory had predicted $T_C \sim 200$ K for $H_3S$ [2], which was corroborated subsequent to the experimental discovery by related work derived from interactions involving both H- and S-like phonons and specifying repulsive Coulomb coupling [3, 8, 10-15];[4] the effects of the vertex correction and electron density of states in *ab initio* calculations are treated in [16]. Some treatments include pressure and structure dependences [2-4, 10, 13, 17]. Owing to its quantum nature, proton dynamics including anharmonic and zero-point motion effects have been treated [4, 18]; calculations reported a second-order S-H bond symmetrization transition, i.e. the quantum phase transition between R3m and $Im\bar{3}m$ structures, occurring at 12 GPa higher pressure for $D_3S$ in comparison to $H_3S$ [4]. The lower $T_C \sim 150$ K for $D_3S$, similar to experimental finding, was also calculated [3, 4, 10]. However, a recent study of the H-D isotope effect indicates a limiting value of the mass isotope exponent $\alpha \leq 0.183$, smaller than the face value $\alpha \approx 0.3$ reported in [1], significantly constraining the degree of possible phonon involvement [9] and thus allowing for consideration of Coulombic mediation. In this work, $H_3S$ is treated as a high-$T_C$ material, in which Coulomb coupling between H and S atoms in adjacent $Im\bar{3}m$ sublattices is considered as a basis for an attractive coupling. Taking the difference in transition temperatures between the $H_3S$ and $D_3S$ samples as arising from variations in disorder, a model formulating $T_C$ derived from earlier research of accepted high-$T_C$ superconductors is presented, yielding an optimal transition temperature $T_{C0} = 198.5$ K at $P = 155$ GPa.

Section 2 presents a brief synopsis of existing $T_C$ data and relevant phononic theoretical predictions. Section 3 presents the Coulomb-based pairing model and its application to $H_3S$. Consideration of the pairing mechanism from the perspective of the Compton scattering of virtual photons, optimal superconductivity of Im$\bar{3}$m H$_3$S in relation to compressed A15 Cs$_3$C$_{60}$ [19], and pair-breaking induced suppression of $T_C$ below $T_{C0}$ for annealing temperatures < RT are discussed in section 4, and the work is summarized in section 5.

---

[3] See online supplemental information for [2] and [5].

[4] See online supplemental information (table V) in [12].

## 2. $T_C$ of H$_3$S

Figure 1 shows experimental results for $T_C$ of H$_3$S over a range of applied pressure, determined from onset points in resistance measurements. Data for samples annealed at or above room temperature ($\geq$ RT), shown as green circles with centered dots, are read from figure 2(c) of [1] and figure 3(c) of [5]; additional data are determined from resistance curves in figures 2(b), 3(b) and 4(a) of [1], figure 3(a) of [5] and figure S3(a) of [6]. For these data, relatively high values of $T_C$ (onsets ~192 – 200 K, midpoints ~190 – 197 K) are found at applied pressures in the range 141 – 155 GPa. Onset $T_C$ data for samples annealed below room temperature (< RT), shown as open blue circles, are taken from figure 1(b) of [1] and illustrate depressed $T_C$ for < RT anneals. Different pressure dependences are also observed; data for $\geq$ RT anneal display a pressure region of maximum $T_C$, whereas $T_C$ for < RT anneal appears monotonic with pressure.

The highest reported onset $T_C$ at 203 K was obtained by extrapolating critical magnetic fields measured at $P$ = 155 GPa for a sample prepared with a $\geq$ RT anneal, where quantum structural calculations predict Im$\bar{3}$m symmetry [4]. The critical fields were determined from points of local magnetization maxima or minima in applied magnetic field sweeps ($\pm$0.2 to $\pm$0.4 Tesla), as reported in figure 4(e) of [1]. Temperature dependence of critical fields $H_{CF}$ and associated magnetization extrema $M_{CF}$ are shown in figure 2, as read from figures 4(c) and (d) in [1], and identified as to the direction of the magnetic field sweep. A notable feature in figure 2(a) is the maximum in $H_{CF}$ at $T$ = 190 – 200 K. The extrapolation $H_{CF} \to 0$ at $T$ = 203.5 K (figure 4(e) in [1]) pertains to data for 200 K $\leq T \leq$ 202.8 K and has been attributed to the signal from the periphery of the sample; this is illustrated in figure 2(b) by the substantially diminished $M_{CF}$ observed for $T \geq$ 200 K [1]. For this sample, the weak field (20 Oe) zero-field-cooled magnetic susceptibility has a transition onset near 200 K (location of the arrow in figure 4(a) of [1]). Accordingly, the highest $T_C$ for H$_3$S is reasonably associated with Im$\bar{3}$m structure, $P$ = 155 GPa and $\geq$ RT anneal.

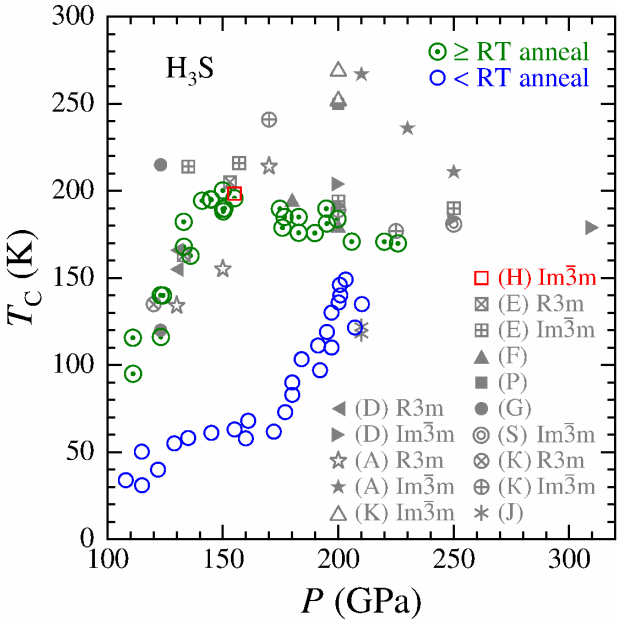

**Figure 1.** Variation of measured $T_C$ of H$_3$S with applied pressure $P$ for $\geq$ RT (room temperature) anneal, green circles with center dot, and < RT anneal, blue circles (from [1, 5, 6]). Gray symbols are various theoretical calculations denoted as (D) [2], (E) [4], (P) [8], (A) [10], (K) [11], (G) [13], (F) [14], (S) [16], (K) [17], and (J) [18]; structure indicated where available. Red square symbol (H) corresponds to $T_{C0}$ from this work.

Currently available theoretical results for $T_C$ of H$_3$S from the phononic theories mentioned in section 1 are represented in figure 1 by various gray symbols [2, 4, 8, 10, 11, 13, 14, 16-18]; the legend indicates the crystal structure where given. In addition to its dependence on pressure and structure, calculated $T_C$ is seen to vary significantly with theoretical methods and assumptions. Calculated values for $T_C$ are larger for the Im$\bar{3}$m structure, when compared to R3m at lower pressure, and range from 122 K determined in [18] up to 267 K in [10] ($P = 210$ GPa taken for both) or 270 K ($P \approx 200$ GPa) in [11]. While these models predict $T_C$ values for both $H_3S$ and $D_3S$ consistent with experiment, they are all essentially based on strong electron-phonon coupling and electron-electron repulsion, and rely on a large H-D isotope effect exponent assumed by taking the difference in measured $T_C$ between these two compounds at face value. However, a recent more detailed analysis of the materials issues involved indicates that much of the observed difference in $T_C$ can be attributed to disorder, particularly evident in $D_3S$, resulting in a reduced value of the isotope effect exponent which is also within error consistent with zero [9]. As a consequence, what initially appeared to be a phonon-mediated material can now be considered unconventional, perhaps high-$T_C$ in nature, where pairing is mediated via Coulombic interactions.

## 3. Coulomb pairing model

Originally formulated for Coulomb interactions between extended layers, the pairing model being considered herein [20] has been shown to be also applicable to local interfacial macrostructures embedded in three-dimensional (3D) extended lattices, as in the case of $Cs_3C_{60}$ [19]. To date, this model has been validated (with a ±1.35 K statistical accuracy) for 50 different layered materials from eight superconducting families (cuprates, ruthenates, rutheno-cuprates, iron-pnictides, BEDT-based [bis(ethylenedithio)tetra-thiafulvalene] organics, respectively in [20-22]; iron-chalcogenides [23]; intercalated group-4-metal nitride-halides [24, 25]; $Cs_3C_{60}$ [19]) with measured $T_C$ values ranging from ~7 to 150 K.

The model [20] assumes an interaction structure comprising two charge reservoirs; a superconducting reservoir, designated as type I, and a mediating reservoir, designated as type II, are physically separated by an interaction distance $\zeta$ defined normal to the layers. In the case of the high-$T_C$ cuprate $La_{1.837}Sr_{0.163}CuO_{4-\delta}$, for example, the interaction occurs between adjacent La/SrO and $CuO_2$ layers, separated by an interaction distance $\zeta$, with (La/Sr)O-(La/Sr)O and $CuO_2$ designated as the type I and type II reservoirs, respectively [20]. For the ostensibly 3D high-$T_C$ organic $Cs_3C_{60}$ (optimized under pressure) the interacting layers are non-planar, where the pairing interaction occurs between the $C_{60}$ molecular macrostructures (type I) and the Cs cations (type II) [19]. Under certain conditions, as are present in the case of $H_3S$, identical reservoirs can function as both types, even in the absence of a conventional layered crystalline structure.

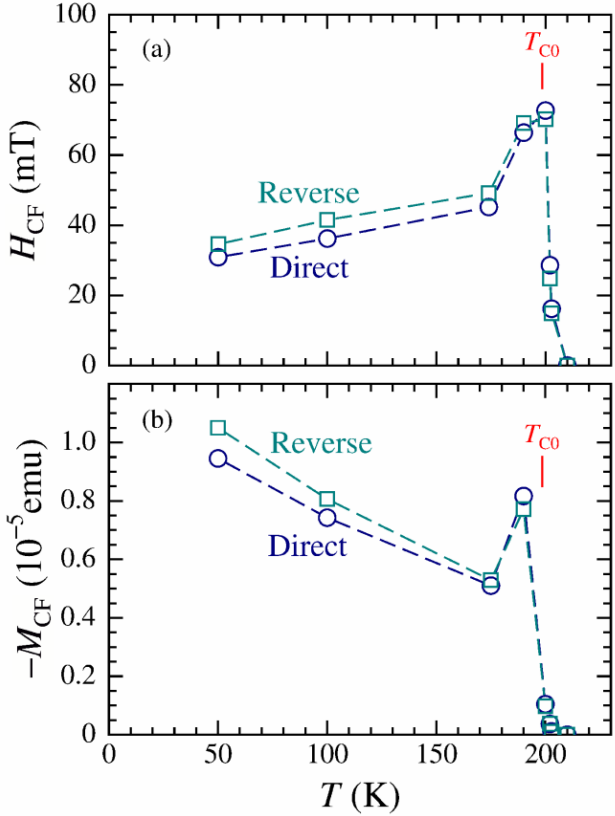

**Figure 2.** Temperature dependence of (a) critical field $H_{CF}$ and (b) corresponding magnetization $M_{CF}$ for $H_3S$ sample at $P = 155$ GPa; labels Direct and Reverse denote directions of magnetic field sweeps (from [1]). Vertical bars mark $T_{C0}$ calculated in this work (section 3.2). The dashed lines are guides to the eye.

## 3.1. $T_{C0}$ and charge allocation

The algebraic expression defining the optimal transition temperature $T_{C0}$ is given by [20],

$$T_{C0} = k_B^{-1} \beta/\ell\zeta = k_B^{-1} (\Lambda/\ell) e^2/\zeta. \qquad (1)$$

The measured constant $\beta = e^2\Lambda = 0.1075 \pm 0.0003$ eV Å$^2$, where $\Lambda = 0.00747(2)$ Å is approximately equal to twice the reduced electron Compton wavelength, was determined previously from experimental data for $T_C$ [20]. The length $\ell = (A/\sigma)^{1/2}$, where $\sigma$ is the participating fractional charge per formula unit and $A$ is the type I basal plane area (per formula unit – equal to the type II area for planar layered structures), relates to the mean spacing between interacting charges. Optimization of the superconducting state is achieved when the two interacting charge reservoirs are in equilibrium; given the systemic materials problems of the high-$T_C$ compounds, which generally act to suppress the measured transition temperature, the value of $\Lambda$ quoted above should be considered a lower limit. The pairing model requires adjacent electronically distinct reservoirs which are present in certain layered structures such as the high-$T_C$ cuprates [20], select 3D macrostructures as in the case of A$_3$C$_{60}$ [19], and manifested in H$_3$S by the two interlaced sublattices of the unit cell.

In evaluating equation (1) for H$_3$S, it is necessary to define and determine the participating fractional charge $\sigma$, the interaction distance $\zeta$, and the area $A$, the latter two being structural properties of the superconducting unit. The fractional charge $\sigma$ associated with an individual charge source, on the other hand, is defined by the general relation,

$$\sigma = \gamma \, v \, [x]. \qquad (2)$$

Here, x represents the available elemental charge and $v$ is the net doping valence; e.g. for La$_{1.837}$Sr$_{0.163}$CuO$_{4-\delta}$, $v_{(La-Sr)} = 3 - 2 = 1$ and $[x]_{(Sr)} = 0.163$ [20]; for Cs$_3$C$_{60}$, $v_{(Cs)} = 1$ and $[x]_{(Cs)} = 3$ [19]. In the former case, doping occurs through ion substitution, whereas the dopant charge in the latter arises from formulaic stoichiometry. The factor $\gamma$ derives from the allocation of the dopant by considering a given compound's structure. Following the procedure applied to high-$T_C$ superconductors, the allocation of charge within the unit cell is determined through rules which define the factor $\gamma$ [20]:

(1a) Sharing between $N$ ions or structural layers/surfaces introduces a factor of $1/N$ in $\gamma$. Examples are $N = 2$ for La$_{1.837}$Sr$_{0.163}$CuO$_{4-\delta}$ and $N = 1$ for both Cs$_3$C$_{60}$ and H$_3$S.

(1b) Doping is shared equally between the two reservoirs, resulting in a factor of 1/2. This rule defines the balance of charge between the two reservoirs required for an optimal superconducting state.

For some optimal compounds (not considered here) $\sigma$ can be calculated by scaling to YBa$_2$Cu$_3$O$_{6.92}$ as discussed in [20-22].

## 3.2. Application to $Im\bar{3}m$ H$_3$S

A recent calculation [2] of the electron localization function (ELF) indicates a value for the S–H bonds of $Im\bar{3}m$ H$_3$S near unity, suggesting a strong polar covalent bond. In contrast, the ELF value for the nearest-neighbor H–H bonds is very low, consistent with the absence of covalent bond characteristics between hydrogen atoms (see figures 6(c) and (d) of [2]). The partial density of states (DOS) for the $Im\bar{3}m$ crystal structure shown in figure 4(b) of [2] and figure 2(b) of [26] indicates an approximately equal distribution of charge (in terms of DOS) between the S and H at the Fermi level. From earlier theoretical work [2], the S-H nearest-neighbor bonds oriented along the three principle axes are determined to be strongly covalent and metallic, whereas the nearest neighbor H-H and next-nearest-neighbor S-H bonds are found to be far weaker. Consequently, the Im$\bar{3}$m unit cell $(H_3S)_2$ structure can be considered as comprising two interlaced simple cubic $H_3S$ metallic and electrostatically-coupled sublattices [8], as is illustrated in figure 3, where each sublattice comprises a three-dimensional network of interconnected ···H–S–H··· chains. The lattice parameter $a_0$ = 3.0823 Å for the highest $T_C$ material observed for $P$ = 155 GPa is determined from the pressure dependence of the atomic volume in figure 2(c) of [5].

Identifying these two interlaced sublattices as the individual reservoirs, and since both the H and S atoms are involved in the superconductivity [2], the Coulomb pairing is mediated through interactions between S atoms in one sublattice and next-nearest-neighbor H atoms in the other. The intersublattice S-H separation is one-half the cube-face diagonal and determines the interaction distance, $\zeta = a_0/2^{1/2}$ = 2.1795 Å (see figure 3). Particularly unique to this structure is that the interaction is between two formula units within the same unit cell, forming the duality where each reservoir serves as both types I and II. For example, in reference to figure 3, the S at cube center interacts with 12 H's at cube edge centers and equivalently an H at a cube face center interacts with 4 S's at neighboring cube corners. Since these interactions occur in three dimensions, the full surface area of the unit cell applies, giving the area per formula unit as $A = 3a_0^2$ = 28.5017 Å$^2$.

Determination of the charge fraction σ assumes a metallic hydrogen model [27], focusing on the charge state of the H atoms from 1s per H as modified by hybridization with S.[5] Integrating the H projected partial density of states calculated for the Im$\bar{3}$m structure in [2] (figure 4(b)) and [26] (figure 2(b)), yields electron charges of 3.33 and 3.53, respectively, with average value 3.43. As in the case of the iron chalcogenides [23], the fractional charge σ in $H_3S$, as defined in equation (1), has two $v$ [x] terms, one from each of the two reservoirs. Assuming the average value above, σ = γ (3.43 + 3.43) = γ (6.86). From rule (1b), this total doping charge is then shared between the two reservoirs such that γ = 1/2, σ = 3.43, ℓ = $(A/\sigma)^{1/2}$ = 2.883 Å, and $T_{C0}$ = 198.5 K ± 3.0 K (taking ± 0.10 uncertainty in $v$ [x]). This result is compared to experimental data and published phononic theoretical predictions in figure 1.

## 4. Discussion

The presence of the Coulomb potential $e^2/\zeta$ in equation (1), together with a length scale approximating the reduced electron Compton wavelength, suggests an interpretation based on superconductive pairing with virtual photons as the mediating bosons. It was previously noted that signatures of the Coulomb potential have been observed in mid-infrared reflectivity of $Cs_3C_{60}$ and thermal reflectance of high-$T_C$ cuprate superconductors [19]. The following discussion considers the origin of the energy $e^2/\zeta$ and the scaling factor $\Lambda/\ell$ contained in the expression for $T_{C0}$ in equation (1), commonalities in the pressure dependences of $T_C$ observed for $H_3S$ and $Cs_3C_{60}$, relation of optical activity in the mid-infrared to $e^2/\zeta$, and association of pair breaking with depressed $T_C$.

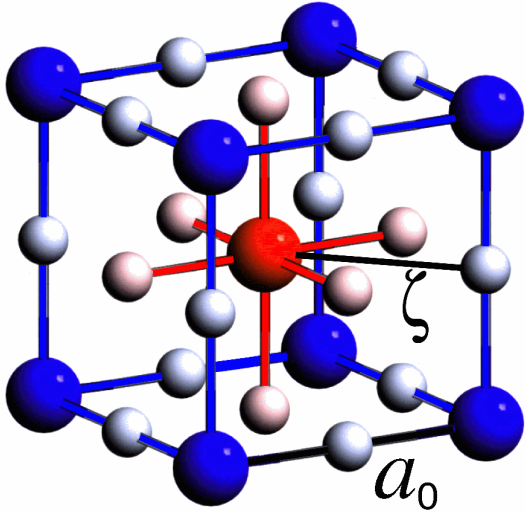

**Figure 3.** Illustration of Im$\bar{3}$m unit cell of compressed $H_3S$ with color contrast distinguishing the two simple cubic sublattices; basis $H_3S$ is shown with S larger than H. Lattice parameter $a_0$ is cube edge; inter-sublattice S-H distance $\zeta$ is one-half cube face diagonal.

---

[5] The H-H bond length calculated for bcc metallic hydrogen in [27] is 1.6% smaller than the S-H bond length $a_0/2$ measured for $H_3S$ at $P$ = 155 GPa [5].

### *4.1 Derivation of $T_{C0}$*

Part of equation (1), the Coulomb potential $e^2/\zeta$ factor, was previously derived (see section 3 of [20]) by drawing an analogy with Allen-Dynes theory [28], which considers BCS superconductivity in the strong-coupling limit, obtaining $T_C \propto (\langle\omega^2\rangle\lambda)^{1/2}$, where $\langle\omega^2\rangle$ is the average square phonon frequency and $\lambda$ the electron-phonon coupling parameter, with Coulomb repulsion being ignored.[6] This derivation evaluates $\langle\omega^2\rangle\lambda$ by an integral over the Fermi surface of the square of electron-ion interaction forces. Evaluated in real space and determined by the inter-reservoir electronic Coulomb interaction (assumed to be unscreened within the unit cell [29]), the force is given as $F(r) = \pm e^2 r\,(r^2 + \zeta^2)^{-3/2}$, which is constrained to the in-plane or longitudinal component for electronic charges in planes separated by a transverse distance $\zeta$ and projected in-plane radial distance $r$. Integrating $2\pi r F^2(r)$ over $r$ from 0 to $\infty$ and taking the square-root yields $(\pi/2)^{1/2} e^2/\zeta$, whence the factor $e^2/\zeta$ in $T_{C0}$ is obtained. In practice, this interaction is local, corresponding to a limited range of integration, $r \leq \zeta$; this results in the same potential factor, but with a smaller numerical coefficient.

The origin of the factor $\Lambda/\ell$ in equation (1) may be understood by considering the Compton scattering probability. Owing to thermal smearing of both the Fermi surface and superconducting gap, one expects ample phase space for quasiparticle creation by Compton scattering events. The essentially instantaneous nature of quasiparticle creation, compared to the significantly slower relaxation process, allows for unscreened Coulomb interactions between electronic charges in adjacent reservoirs. Equation (1) thus derives fundamentally from quantum fluctuations within the virtual photon field between the two charge reservoirs inducing Compton scattering of quasiparticles confined essentially within their respective reservoirs.

Scattering of a quasiparticle from an initial equilibrium state $\boldsymbol{k}$ on the Fermi surface to a final state $\boldsymbol{k'}$ via a Compton scattered virtual photon precipitates a corresponding shift from $\boldsymbol{q}$ to $\boldsymbol{q'}$ in the photon wavevector. The final state wavevector $\boldsymbol{k'}$ is expected to contain a small transverse component, such that $\boldsymbol{k}$ and $\boldsymbol{k'}$ remain essentially transverse to the direction of the separation distance $\zeta$. Such electron-photon scattering events can be likened to the problem of electron-impurity scattering, where transport electrons confined to near the Fermi surface with Fermi wavevector $k_F$ are scattered by real space localized potentials with a mean free path spacing of $\ell_{mfp}$; the probability for a single such scattering event scales as $(k_F \ell_{mfp})^{-1}$. In the Compton scattering picture, virtual photons of wave vector $\boldsymbol{q}$ confined within the real space region between the two interacting reservoirs defined by $\zeta$ act in analogy to transport electrons and scatter off of the essentially localized electrons/holes of spacing $\ell$ (acting in analogy to impurities). The analogy entails a tenable difference between electron and photon time scales, given Fermi velocity $v_F \ll c$. Thus, the probability of scattering a virtual photon from initial state $\boldsymbol{q}$ to final state $\boldsymbol{q'}$ is given by the differential $(q'\ell)^{-1} - (q\ell)^{-1} = \Delta\lambda \ell^{-1} = \lambdabar_C(1 - \cos\theta)\ell^{-1}$, where $\lambdabar_C = \hbar/m_e c$ is the reduced Compton wavelength and $\theta$ is the angle between the incident and scattered photons (note that setting $\theta = \pi$ yields $\Delta\lambda = 2\lambdabar_C$). The momentum perturbation associated with electron recoil is very small, since $|\boldsymbol{q} - \boldsymbol{q'}| \ll k_F$ for virtual photons of energy $e^2/\zeta$. Identifying $\Lambda$ as proportional to $\lambdabar_C$ and integrating over the relevant angles of the incident and scattered photons one obtains,

$$T_{C0} \propto (\lambdabar_C/\ell)\, e^2/\zeta, \qquad (3)$$

---

[6] See e.g. equation (25) in [28]

a result which can be (holistically) interpreted as the product of the scattering probability and energy of the scattering interaction. Thus, the mediation of the high-$T_C$ superconducting state is facilitated through Compton scattering between virtual photons and unscreened quasiparticles.

### *4.2 Optimal compression in $H_3S$ and $Cs_3C_{60}$*

The maximum in the pressure dependence of $T_C$ for $H_3S$ resembles that for compressed $Cs_3C_{60}$, where the pressure for maximum $T_C$ has been identified as being optimal for high-$T_C$ Coulombic coupling [19]. Proposing such a comparison recognizes that the R3m-Im$\bar{3}$m transition in $H_3S$ at 140 GPa [7] occurs in a flat, pressure-independent region of $T_C(P)$ (see figure 1). The $T_C$ versus $P$ diagrams for both materials are bounded at low pressure by semiconducting-like phases [1, 30], whereas superconductivity persists at high pressures, although with diminished $T_C$. Optimization of high-$T_C$ superconductivity in $H_3S$ and $Cs_3C_{60}$ with applied pressure is directly analogous to the behavior observed in $YBa_2Cu_4O_8$, where maximum $T_C$ is achieved for $P \approx 12$ GPa [31]. The dome-like dependency of $T_C$ on $P$ is attributed, in that case, to charge transfer between the two reservoirs with the peak in $T_C$ corresponding to an optimal, equilibrium state [20].

The suppression of $T_C$ in $H_3S$ with decreasing applied pressure can be shown to be attributable to the concomitant increase in rhombohedral distortion detected in x-ray diffraction below about 140 GPa [7], which appears to be primarily associated with the R3m phase forming two different H-S bond lengths and off-center hydrogens [2, 4]. Specifically, a splitting in several peaks is observed in [7], which is also reflected in the presumably lower-resolution data of [5] exhibiting an increased broadening with decreasing pressure. It is likely that the disorder introduced by the rhombohedral distortion and asymmetric H-S bonding act to suppress $T_C$ below the optimal level. It is also quite likely that inhomogeneities artificially broaden the R3m/Im$\bar{3}$m transition, with the peak in $T_C(P)$ corresponding to the pressure at which a largely homogeneous Im$\bar{3}$m phase is achieved.

### *4.3 Optical activity in the mid-infrared*

There are also local structural similarities between Im$\bar{3}$m $H_3S$ and A15 $Cs_3C_{60}$. A given S in $H_3S$ has 12 H next-nearest-neighbors in the other sublattice, forming rhombic dodecahedral symmetry with an H over each of the 12 faces. In A15 $Cs_3C_{60}$, a given $C_{60}$ also has 12 Cs nearest neighbors located at the 12 vertices of a regular convex icosahedron. In the case of A15 $Cs_3C_{60}$ near optimal compression, a maximum near ~2500 cm$^{-1}$ (0.31 eV) in the mid-infrared optical conductivity at ambient temperature [32] (figure 1I therein) was previously identified with an excitation scaling with the energy $e^2/\zeta$ [19]. For $H_3S$ at $P = 150$ GPa, the normal-state reflectance shows a depressed band with minimum at ~0.46 eV [6]. The respective energies of these mid-infrared signatures are in the ratio ~0.674. One notes in comparing $\zeta = 3.1969$ Å for $Cs_3C_{60}$ to $\zeta = 2.1795$Å for $H_3S$ that the values of $\zeta^{-1}$ are in the ratio 0.676. The fact that these two ratios are essentially the same is expected for boson mediators scaling with $e^2/\zeta$, assuming equivalent excitations configurations and optical dielectric constants.

### *4.4 Pair-breaking depression of $T_C$*

Contained in the various resistivity versus temperature traces $R(T)$ presented in [1], with independent variables of applied pressure and sample thermal annealing associated with cycling the temperature, is the functional dependence of $T_C$ on resistance for the $H_3S$ samples. Of particular relevance are electron scattering rates associated with sample impurities and defects, as determined from the residual resistance $R(0)$. Values of $R(0)$ are obtained by linear extrapolations of $R(T)$ data in the normal-state region $T > T_C$. Using the onset of the resistance transition as the consistent measure of $T_C$, the $T_C$-$R(0)$ data obtained from figures 1(a), 2(a) and 2(b) of [1] are plotted in figure 4 and distinguished as to annealing procedures, similarly as is done in figure 1. An additional datum with $T_C = 200.3$ K determined from figure S3(a) in [6] is plotted at $R(0) = 0$. As noted in [1], $R(T)$ exhibits semiconductor-like temperature dependences in the normal state (i.e. $dR/dT < 0$) for pressures below 135 GPa. Data estimation in this region yields $R(0) \approx 9.6, 26, 29,$ and $65\ \Omega$, and $T_C \approx 54, 33, 32,$ and 23 K, corresponding to $P = 129, 122, 115,$ and 107 GPa, respectively, producing points lying off scale to the right in figure 4. There is a change in $T_C$ trend near $R(0) \sim 1\ \Omega$ ($P \sim 177 - 192$ GPa), reflecting the change in trend of $T_C$ versus $P$ (figure 1) that has been attributed to a phase transformation [1]. The data for the region $R(0) < 1\ \Omega$, corresponding to $T_C \geq 89$ K and $P \geq 192$ GPa, show a systematic and nearly linear decrease in $T_C$ with increasing $R(0)$ and are assumed to correspond to Im$\bar{3}$m phase H$_3$S material.

The strong correlation between $T_C$ and $R(0)$ points to the extrinsic dependence of $T_C$ on defect scattering such as to induce disorder-related pair-breaking [21]. Taking the highest measured resistive onset of $T_C = 200.3$ K from [6] as a fixed point $T_C^{max}$, the observed transition temperature $T_C$ is thus modeled by the pair-breaking expression,

$$T_C = T_C^{max} \exp[\psi(\tfrac{1}{2}) - \psi(\tfrac{1}{2} + \alpha_p/2\pi k_B T_C)], \quad (4)$$

where $\alpha_p$ is the pair-breaking parameter, $k_B$ is Boltzmann's constant, and $\psi$ is the digamma function [33]. Values of $\alpha_p$ were determined from equation (4) for each measurement of $T_C$. For $R(0) < 1\ \Omega$, the results closely follow the linear relationship $\alpha_p = s\,R(0)$, with scaling coefficient $s = 12.2 \pm 0.4$ meV/$\Omega$ and coefficient of determination $R^2 = 0.975$. The dashed curve in figure 4 is calculated from equation (4) with $\alpha_p = s\,R(0)$.

The pair-breaking rate $\alpha_p/2\hbar$ may be compared to the electron-defect scattering rate by invoking the free-carrier expression $\tau(0)^{-1} = ne^2 R(0) t/m^*$, in terms of carrier density $n$, sample thickness $t$, and effective mass $m^*$. Taking the estimated thickness $t \sim 1\ \mu$m and determining $ne^2/m^*$ from the London penetration depth $\lambda_L = 125$ nm, both as given in [1],[7] one obtains $\hbar\tau(0)^{-1} \approx (0.42\ \text{eV}/\Omega)\,R(0)$. For example, a sample $\geq$ RT annealed at $P = 145$ GPa with $T_C = 195.0$ K has estimated residual scattering $\hbar\tau(0)^{-1} \approx 15$ meV and calculated $\alpha_p = 0.55$ meV. Somewhat larger residual scatterings were determined from non-linear fits to $R(T)$ in [6]. Thus, while the pair-breaking rate is a relatively small fraction ($< 2\%$) of the defect scattering rate, the resulting effect is nevertheless strong enough to induce depressions in $T_C$ by up to 50% relative to $T_{C0}$ for $R(0)t$ up to $\sim 0.1$ m$\Omega$cm.

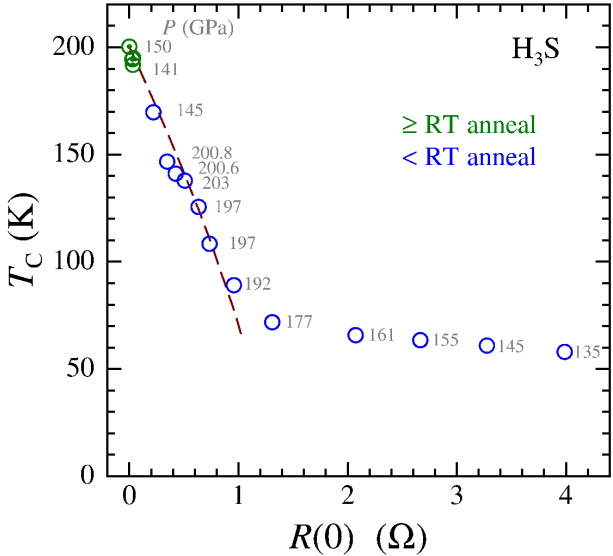

**Figure 4.** Variation of measured $T_C$ of H$_3$S for annealing temperatures $\geq$ RT (green circles with center dot) and $<$ RT anneal (blue circles) versus residual electrical resistance $R(0)$; corresponding applied pressures $P$ are indicated adjacent to the data points (from [1] and [6]). The dashed curve represents the fitted Eq. (4) with $\alpha_p = s\,R(0)$.

---

[7] Alternative analyses presented in [34] yield $\lambda_L = 163 - 189$ nm.

## 5. Conclusion

Given the conflict between the actual isotope effect exponent α ≤ 0.183 with the assumptions made in existing phonon-based theories, along with other inconsistent behaviors exhibited in the experimental data, it is evident that a purely phononic approach falls somewhat short of explaining the superconductivity in $H_3S$. In the present work, a Coulomb mediation model is instead proposed in which the pairing interaction occurs between the two interlaced simple cubic $H_3S$ metallic sublattices depicted in figure 3. This is possible since calculations of S-H bonding in the $Im\bar{3}m$ structure show the nearest-neighbor bonds along the principle cube axes to be strongly covalent and metallic, while the next-nearest-neighbors of S-H form weak bonds [2]. Unlike previous compounds studied, both the type I and type II reservoirs are identical, with the interaction occurring across the S-H next-nearest-neighbors. Evaluation of equation (1) gives $T_{C0} = 198.5 \pm 3.0$ K at $P = 155$ GPa, which compares well with resistivity onset, magnetic susceptibility onset and critical field data (the average is 200.0 K). From this analysis, it appears that interactions between quasi-isolated H-M-H linear structures are essential for the production of superconductivity in the $H_3M$ system. The observed suppression of $T_C$ below $T_{C0}$ with increasing residual resistivity $R(0)$ in the $Im\bar{3}m$ phase is then understood from the perspective of scattering-induced pair breaking.

The origin of equation (1) is also considered from the perspective of quantum fluctuations in the inter-reservoir virtual photon field, where quasiparticle excitation arises via Compton scattering. Owing to the near-instantaneous quasiparticle creation, compared to the slower relaxation time, the Coulomb interaction between charges in adjacent reservoirs is essentially unscreened. Drawing on an analogy with the Allen-Dynes equation for strong coupling in which the interaction is the inter-reservoir Coulomb potential, and positing that $T_{C0}$ is also proportional to the virtual photon differential scattering probability, the two principal factors $e^2/\zeta$ and $\lambda_C/\ell$ comprising equation (1) are respectively derived. Experimental confirmation of the $e^2/\zeta$ energy scale is found through a comparative analysis with $Cs_3C_{60}$ of mid-infrared reflectivity data showing correlation between energy and $\zeta^{-1}$.

In Coulomb-mediated (high-$T_C$) superconductors, phononic interactions have been generally observed to be associated with pair breaking [19]. While the model presented herein fully explains the superconductivity in $H_3S$ and accurately determines $T_{C0}$, experimental uncertainty in α leaves open the possibility of a hybrid mechanism incorporating Coulombic enhancements to a BCS-based model. Additionally, understanding the material yielding the 203.5-K extrapolated transition continues to be of interest. Remaining to be determined is whether forms of indirect Coulomb or excitonic coupling [35], analogous to that proposed for $H_3S$ in this work, act to enhance high-$T_C$ superconductivity in other highly compressed hydride materials under consideration, e.g. in [36] and reviewed in [37, 38]. In searching for promising high-$T_C$ compounds, the formula of equation (1) suggests such materials comprise the following attributes: (1) substructures distinguished by spatial separation and bonding, (2) at least one of the substructures contains the type I charge reservoir with extended metallic electron states, (3) high interaction fractional charge σ from charge transfer or doping, (4) short interaction distance $\zeta$ between the substructures, and (5) small value of formulaic area $A$. Applying these five requirements for the formation of a high-$T_C$ superconducting state to computational prediction techniques for new high-$T_C$ materials is a subject for future work.

## Acknowledgements

The authors are grateful for support from the College of William and Mary, New Jersey Institute of Technology, and University of Notre

Dame, and A. F. Goncharov, private communication.

**References**

[1] Drozdov A P, Eremets M I, Troyan I A, Ksenofontov V and Shylin S I 2015 *Nature* **525** 73

[2] Duan D, Liu Y, Tian F, Li D, Huang X, Zhao Z, Yu H, Liu B, Tian W and Cui T 2014 *Sci. Rep.* **4** 6968

[3] Errea I, Calandra M, Pickard C J, Nelson J, Needs R J, Li Y, Liu H, Zhang Y, Ma Y and Mauri F 2015 *Phys. Rev. Lett.* **114** 157004

[4] Errea I, Calandra M, Pickard C J, Nelson J R, Needs R J, Li Y, Liu H, Zhang Y, Ma Y and Mauri F 2016 *Nature* **532** 81

[5] Einaga M, Sakata M, Ishikawa T, Shimizu K, Eremets M I, Drozdov A P, Troyan I A, Hirao N and Ohishi Y 2016 *Nat. Phys.* **12** 835

[6] Capitani F, Langerome B, Brubach J-B, Roy P, Drozdov A, Eremets M I, Nicol E J, Carbotte J P and Timusk T 2017 *Nat. Phys.* **13** 589

[7] Goncharov A F, Lobanov S S, Prakapenka V B and Greenberg E 2017 *Phys. Rev. B* **95** 140101(R)

[8] Papaconstantopoulos D A, Klein B M, Mehl M J and Pickett W E 2015 *Phys. Rev. B* **91** 184511

[9] Harshman D R and Fiory A T 2017 *Supercond. Sci. Technol.* **30** 045011

[10] Akashi R, Kawamura M, Tsuneyuki S, Nomura Y and Arita R 2015 *Phys. Rev. B* **91** 224513

[11] Komelj M and Krakauer H 2015 *Phys. Rev. B* **92** 205125

[12] Bernstein N, Hellberg C S, Johannes M D, Mazin I I and Mehl M J 2015 *Phys. Rev. B* **91** 060511(R)

[13] Gor'kov L P and Kresin V Z 2016 *Sci. Rep.* **6** 25608

[14] Flores-Livas J A, Sanna A and Gross E K U 2016 *Eur. Phys. J. B* **89** 63

[15] Szczęśniak R and Durajski A P 2017 *Solid State Commun.* **249** 30

[16] Sano W, Koretsune T, Tadano T, Akashi R and Arita R 2016 *Phys. Rev. B* **93** 094525

[17] Kudryashov N A, Kutukov A A and Mazur E A 2017 *Nov. Supercond. Mater.* **3** 1

[18] Jarlborg T and Bianconi A 2016 *Sci. Rep.* **6** 24816

[19] Harshman D R and Fiory A T 2017 *J. Phys.: Condens. Matter* **29** 145602

[20] Harshman D R, Fiory A T and Dow J D 2011 *J. Phys.: Condens. Matter* **23** 295701

Harshman D R, Fiory A T and Dow J D 2011 *J. Phys.: Condens. Matter* **23** 349501 (corrigendum)

[21] Harshman D R and Fiory A T 2012 *Phys. Rev. B* **86** 144533

[22] Harshman D R and Fiory A T 2015 *J. Phys. Chem. Sol.* **8** 106

[23] Harshman D R and Fiory A T 2012 *J. Phys.: Condens. Matter* **24** 135701

[24] Harshman D R and Fiory A T 2014 *Phys. Rev. B* **90** 186501

[25] Harshman D R and Fiory A T 2015 *J. Supercond. Nov. Magn.* **28** 2967

[26] Oh H, Coh S and Cohen M L 2016 *arXiv*:1606.09477v2 [cond-mat.supr-con]

[27] Wigner E and Huntington H B 1935 *J. Chem. Phys.* **3** 764

[28] Allen P B and Dynes R C 1975 *Phys. Rev. B* **12** 905

[29] Varma C M, Schmitt-Rink S and Abrahams E 1987 *Solid State Commun.* **62** 681

[30] Ganin A Y, Takabayashi Y, Jeglič P, Arčon D, Potočnik A, Baker P J, Ohishi Y, McDonald M T, Tzirakis M D, McLennan A, Darling G R, Takata M, Rosseinsky M J and Prassides K 2010 *Nature* **466** 221

[31] Scholtz J J, van Eenige E N, Wijngaarden R J and Gressen R 1992 *Phys. Rev. B* **45** 3077

[32] Baldassarre L, Perucchi A, Mitrano M, Nicoletti D, Marini C, Pontiroli D, Mazzani M, Aramini M, Riccó M, Giovannetti G, Capone M and Lupi S 2015 *Sci. Rep.* **5** 15240

[33] Tinkham M 1996 *Introduction to Superconductivity*, 2nd edn (NewYork: McGraw Hill)

[34] Talantsev E F, Crump W P, Storey J G and Tallon J L 2017 *Ann. Phys.* (Berlin) **529** 1600390

[35] Little W A 1987 *Novel Superconductivity* ed S A Wolf and V Z Kresin (NewYork: Plenum) p 341

[36] Fu Y, Du X, Zhang L, Peng F, Zhang M, Pickard C J, Needs, R J, Singh D J, Zheng W and Ma Y 2016 *Chem. Mater.* **28** 1746

[37] Duan D, Liu Y, Ma Y, Shao Z, Liu B and Cui T 2017 *Natl. Sci. Rev.* **4** 121

[38] Shamp A and Zurek E 2017 *Nov. Supercond. Mater.* **3** 14